# Mechanical alloying of Cu and Fe induced by severe plastic deformation of a Cu-Fe composite.

X. Sauvage[1], F. Wetscher[2], P. Pareige[1]

[1] Groupe de Physique des Matériaux - UMR CNRS 6634, Institut of Material Research, Université de Rouen, 76801 Saint-Etienne-du-Rouvray, France.

[2] Erich Schmid Institute of Material Sciences, CD-Laboratory for Local Analysis of Deformation and Fracture, Austrian Academy of Sciences, Jahnstraße 12, A-8700 Leoben, Austria

*Abstract*

A filamentary composite elaborated by cold drawing was processed by High Pressure Torsion (HPT). The nanostructure resulting from this severe plastic deformation (SPD) was investigated thanks to scanning electron microscopy, transmission electron microscopy, X-ray diffraction and 3D atom probe. Although the mutual solubility of Cu and Fe is extremely low at room temperature in equilibrium conditions, it is shown that nanoscaled Fe clusters dissolve in the Cu matrix. The non-equilibrium copper supersaturated solid solutions contain up to 20at.% Fe. The driving force of the dissolution is attributed to capillary pressures and mechanisms which could enhanced the atomic mobility during HPT are discussed. We conclude that the interdiffusion is the result of a dramatic increase of the vacancy concentration during SPD.

1. **Introduction**

The microstructure of alloys processed by severe plastic deformation (SPD) has been widely investigated during the past two decades. Many different techniques have been developed such as Equal Channel Angular Extrusion (ECAE) [1-3], High Pressure Torsion (HPT) [1,4-11] and Accumulated Roll Bonding (ARB) [12] to produce nanostructured metallic materials. The microstructure evolution of pure metals under SPD is well documented but much less is known and reported about multi-phase materials. However, previous investigations of cold rolled laminates [13-16] and cold drawn metal matrix composites [17-20] have shown that multi-phase or multi-material mixtures could be excellent candidates for the achievement of nanostructures with unique properties.

During severe plastic deformation, a large amount of defects is created : the dislocation density as well as the vacancy concentration dramatically increase and new high angle grain boundaries are formed [1]. All these defects, along with internal stresses resulting from the plastic deformation [20], may promote phase transformations such as disordering [21], solid state amorphization [15,22] or dissolution of precipitates [2,3,23,24]. Similar features have been reported in powder mixtures processed by ball milling. In that case, friction and strong impacts of steel balls on powder grains could induce the alloying of non-miscible elements [16,25-27].

The aim of this work was to produce by SPD a non-equilibrium solid solution of two non-miscible elements in a bulk specimen. The Fe-Cu system was chosen because it has been widely investigated [13,16,18,20,25,26,28].

Following the work of Yavari and co-authors [16], the mechanical alloying of Fe in a Cu matrix could simply be explained by thermodynamical arguments (Gibbs-Thomson effect) : intense deformation would generate nanometer scaled Fe fragments which are unstable because of capillary pressures. In order to produce such a nanoscaled structure and to initiate the alloying process, slices of a Cu-Fe filamentary composite elaborated by cold drawing were deformed by High Pressure and Torsion (HPT). In this paper we discuss the microstructure resulting from SPD and the possible mechanisms of the Fe-Cu interdiffusion during HPT.

## 2. Experimental procedure

### 2.1 Fabrication of the Cu-Fe composite

The initial Cu-Fe composite material was produced by drawing together Cu (99.5% purity) and Fe (99.5% purity) components. The technique used is a variant of the process invented by Levi [18], and recently described by Thilly and co-authors [17]. At first, a copper tube (diameter 12mm, thickness 1mm) was filled with an iron tube (diameter 8.5mm, thickness 1.5mm) which was filled itself with a copper wire (diameter 5mm). This rod was cold drawn down to 1mm in diameter. Subsequently, the wire was cut in 51 short pieces that were inserted into another copper tube (diameter 12mm, thickness 1mm). This rod was then cold drawn down to 1mm in diameter. At the third step, the same procedure was repeated. Finally, the composite was cut in 72 short pieces that were inserted into another copper tube (diameter 14mm, thickness 1mm) and then cold drawn down to 8mm. The volume fraction of Fe is about 10% in the inner part of the composite (without taking into account the last outer Cu tube). Tubes and wires were carefully cleaned before assembling and annealed 60 min at 923K as soon as 90% of area reduction was reached, in order to avoid fractures.

The microstructure of the Cu-Fe composite exhibits an homogenous distribution of Fe filaments in the copper matrix (figure 1). The thickness of these filaments is about a few micrometers.

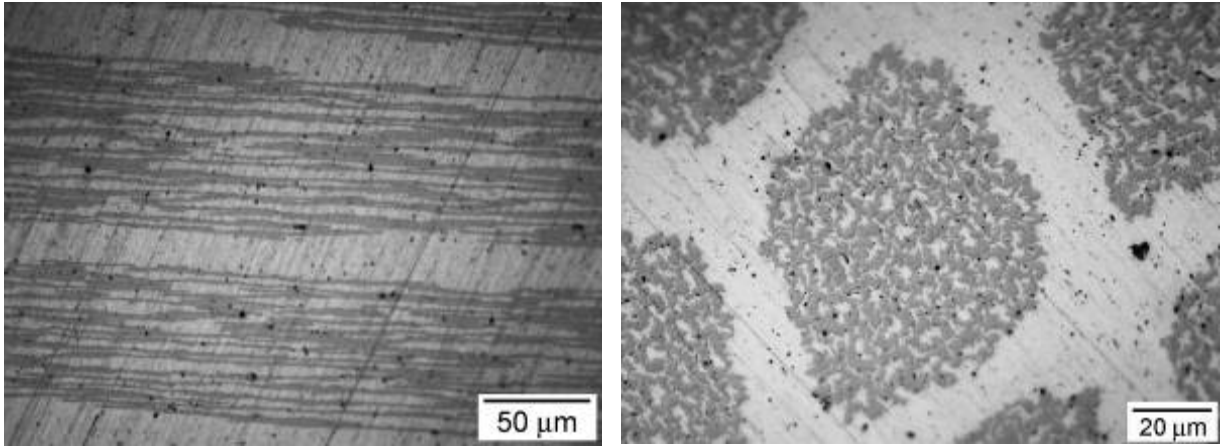

a)                                           b)

*Figure 1 : Optical micrographs of the Cu-Fe composite wire fabricated by cold drawing. Cu is imaged in bright and Fe in dark. (longitudinal view (a) and cross section view (b)).*

## 2.2 Deformation of the Cu-Fe composite by HPT

Thin slices of the Cu-Fe composite (diameter $d_s$=8mm, initial thickness $t_i$=0.6mm) were processed by High Pressure and Torsion (pressure applied P=5 GPa, N=5 turns of torsion and rotation speed 0.2 turn/min) with the wire axis of the composite parallel to the torsion axis. Due to the compression, the final thickness of the sample was approximately $t_s$=0.3mm.

During the deformation by torsion, if it is assumed that Fe filaments are continuously sheared, they should become thin layers with a thickness $t_L$ that could be estimated considering the volume conservation of the filament. The expression of $t_L$ is given by :

$$t_L \approx t_s\, d_f / 2\Pi N r \qquad (1)$$

where r is the distance from the centre of the disc to the filament and $d_f$ the average diameter of the filament. Thus, a 1μm thick filament ($d_f = 10^{-6}$ m) at a distance of 2mm from the centre (r = 2 $10^{-3}$ m) should become a 5nm thick layer after 5 turns. It is though that such a nanostructure could initiate the alloying of Fe and Cu [16].

## 2.3 Investigation of the microstructure

Investigations were carried out by optical microscopy (Olympus BX51M) and scanning electron microscopy (LEO FE1530, field emission gun, back scattered electron detector).

Transmission Electron Microscopy observations were performed with a JEOL 2000FX microscope operating at 200 kV. TEM samples were prepared by ion milling (GATAN DUOMILL 600DIF, 5kV, angle 12°, 300K). Samples with a diameter of 3mm were cut in the discs processed by HPT at a distance of 2±0.5 mm from the disc center. Thus, TEM investigations were performed with the electron beam parallel to the torsion axis.

X-ray diffraction (XRD) patterns were taken with a Brucker D8 system in Bragg-Brentano θ-2θ geometry. The X-ray generator was equipped with a Co anticathode, using Co (Kα) radiation (λ = 0.17909 nm). The measurements were performed so that the out-of-plane component was the torsion axis of the sample processed by HPT.

Field Ion Microscopy (FIM) images were obtain at 50K using Ne as imaging gas (pressure of $5 \times 10^{-3}$ Pa). 3D atom probe (3D-AP) analysis were carried out at 50K in UHV (residual pressure $10^{-8}$ Pa), with 17% pulse fraction and 2kHz pulse repetition rate. The position sensitive detector used was a CAMECA's Tomographic Atom Probe (TAP) detector. FIM and 3D-AP specimens were prepared by electropolishing (10g $Na_2CrO_4(H_2O)_4$ + 100ml acetic acid, 8V, 300K). Small rods were cut in HPT discs and needle shaped specimens were prepared so that the tip was located at a distance of 2±0.5 mm from the disc center (see reference [24] for details).

## 3. Experimental results

### *3.1 Microstructure of the Cu-Fe composite prior to HPT deformation*

Prior to HPT deformation, the Cu-Fe composite exhibits continuous Fe filaments elongated along the wire axis (figure 1). The thickness of the original Fe tubes has decreased down to a few micrometers during the drawing process. However they are curled in the cross section and do not exhibit a regular cylindrical shape in the cross section. This feature results from the non-axisymmetric deformation of the α-Fe filaments which is due to the strong (110) drawing texture of this BCC phase [30]. This texture is clearly revealed by the diffraction pattern (figure 2-a) where only the (110) peak of the α-Fe phase appears (the experiment was performed so that the out-of-plane component was parallel to the wire axis). A typical (111) drawing texture of the FCC copper phase is also exhibited.

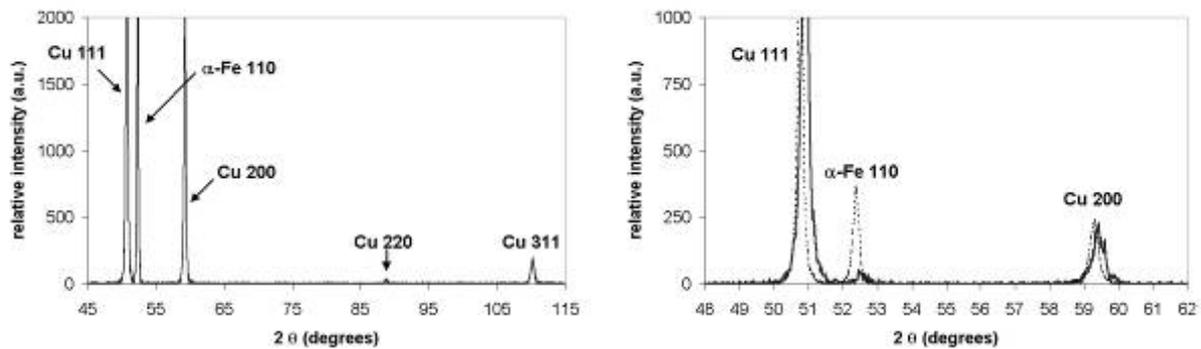

a)                                                                                      b)
*Figure 2 : (a) X-ray diffraction patterns of the Cu-Fe composite prior to HPT deformation. (b) Comparison between the diffraction pattern prior to HPT deformation (dashed line) and after 5 turns.*

## 3.2 Microstructure after 5 turns of HPT

The HPT deformation induces a slight broadening of diffraction peaks (figure 2-b) which might be attributed to internal stresses and to the reduction of the mean grain size. The α-Fe peak is strongly attenuated indicating that this phase might have partly vanished or that nanometer scaled α-Fe particle have been created during the HPT deformation. The slight shift of the FCC copper peaks could indicate a decrease of the lattice parameter. This feature would be discuss later in this paper.

As expected, the microstructure in the cross section of the samples exhibits a layered structure (figure 3-a). The contrast of back scattered electron SEM images results from the higher scattering factor of the Cu phase. This indicates that Cu and Fe are not homogeneously distributed. High magnification images show a very fine scaled structure (figure 3-b) : some Fe rich layers (black contrast) are only 10 nm thick.

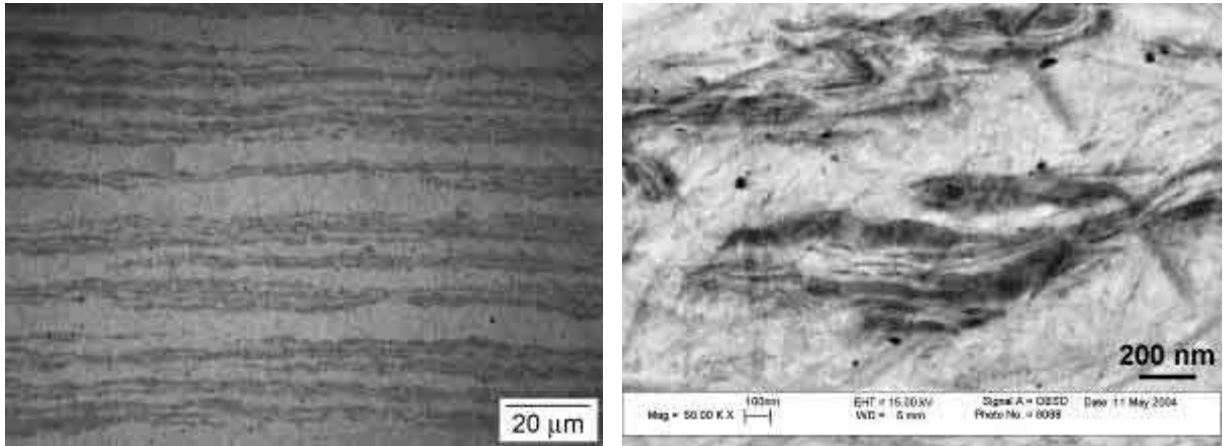

*Figure 3 : Layered microstructure of the Cu-Fe composite processed by HPT after 5 turns (the torsion axis is parallel to the vertical direction of the pictures) ; a) optical micrograph x100, Cu is imaged in bright ; b) SEM back scattered picture x 50k, Cu is imaged in bright.*

The TEM pictures (figure 4) do not show the layered structure because the foils were prepared so that the electron beam was parallel to the torsion axis (i.e. perpendicular to the layers). Some regions are predominantly formed of FCC copper grains with a grain size in a range of 100 to 200 nm (figures 4-a and 4-b). Other regions are a mixture of both FCC copper and BCC $\alpha$-Fe grains with a smaller grain size in a range of 20 to 80 nm (figures 4-c and 4-d).

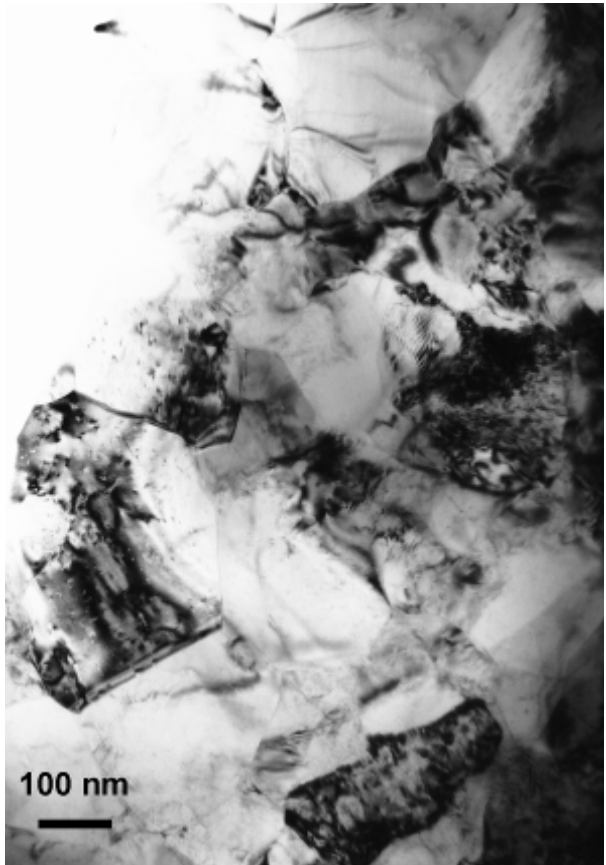
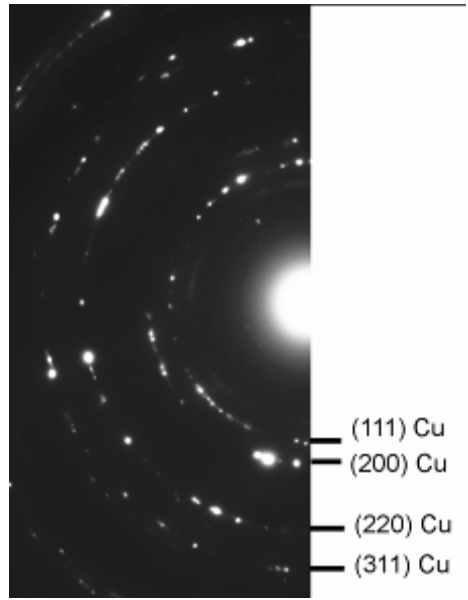

a)  
b)

(111) Cu
(200) Cu
(220) Cu
(311) Cu

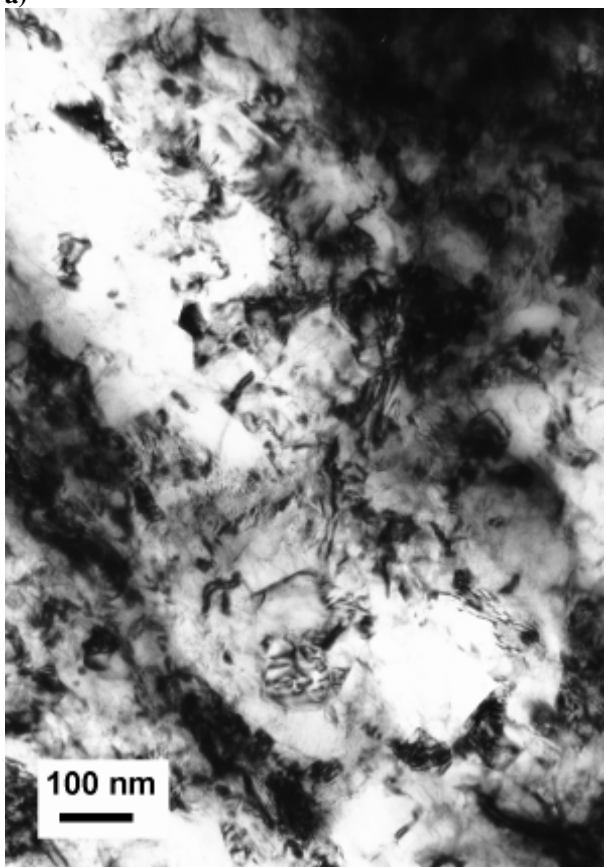
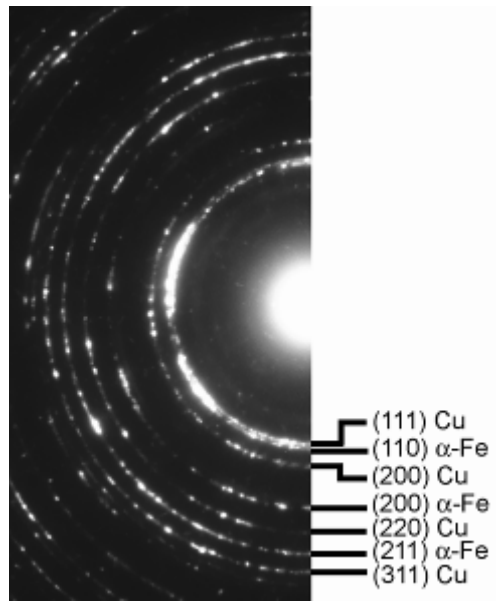

c)  
d)

(111) Cu
(110) α-Fe
(200) Cu
(200) α-Fe
(220) Cu
(211) α-Fe
(311) Cu

*Figure 4 : TEM bright field images of the Cu-Fe composite after 5 turns by HPT, a) α-Fe free area and related SAED pattern (b) ; c) Cu and α-Fe mixed area and related SAED pattern (d).*

## 3.3 FIM and 3D Atom Probe investigation of the nanostructure

FIM pictures (figure 5) show a strong contrast resulting from the higher evaporation field of Fe atoms [31]. Fe rich grains are brightly imaged because the electric field is locally higher leading to a higher ionisation rate of Ne gas atoms. FIM images exhibit very small Fe clusters with a grain size in a range of 5 to 50 nm. These nanograins probably result from the shearing of the original Fe filaments of the Cu-Fe composite. As already pointed out by SEM observations, it appears that Cu and Fe atoms are not homogeneously distributed after the HPT process.

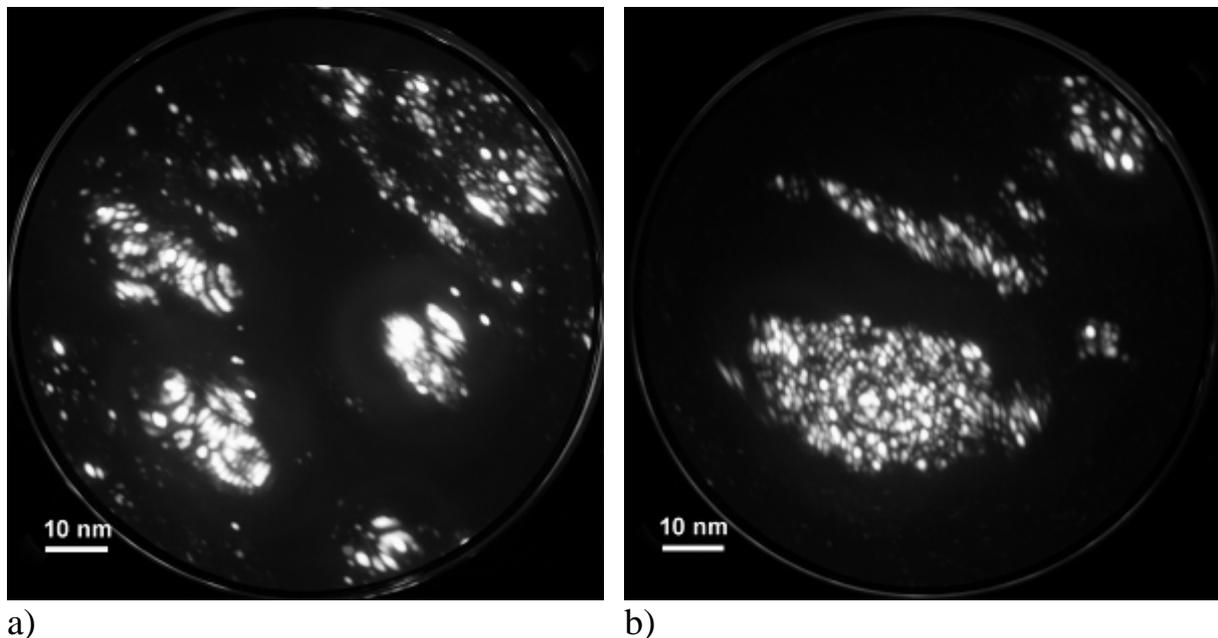

a) b)

*Figure 5 : Field ion microscopy images of the Cu-Fe composite after 5 turns by HPT, the brightly imaged zones are $\alpha$-Fe .*

3D-AP analysis were performed to map out the distribution of Cu and Fe atoms. Figure 6-a shows the 3D distribution of Cu atoms in the vicinity of a Fe cluster which diameter is about 10 nm. The Fe concentration in the copper phase is 4at.%±0.5 while the amount of Cu detected in the Fe cluster is not significant regarding the error bar. The Cu concentration profile (figure 6-b) exhibits a sharp gradient at the interface. This 1 nm wide gradient is attributed to the roughness of the interface and to the thickness of the sampling volume used to plot the profile.

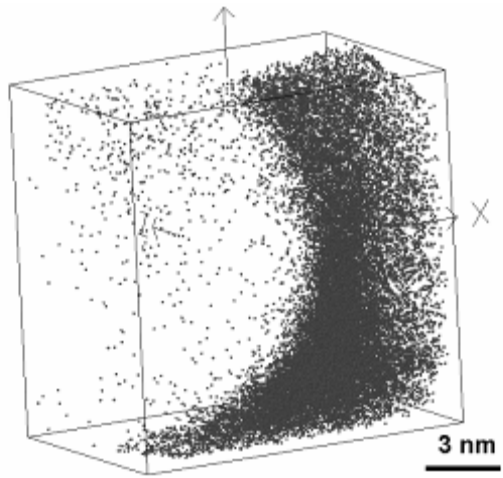
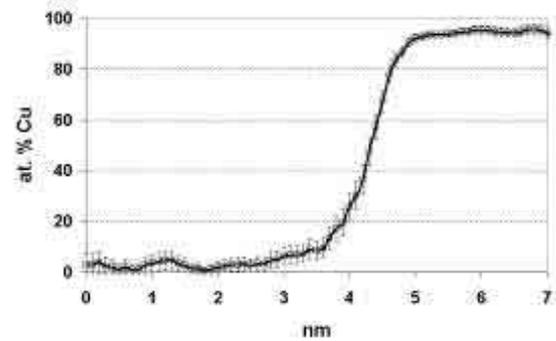

a)                                          b)

*Figure 6 : 3D-AP data set of the Cu-Fe composite after 5 turns by HPT, a) 3D reconstructed volume (12x12x8 nm³), only Cu atoms are displayed to exhibit the interface between a Fe cluster and the Cu matrix, b) Copper concentration profile across the Cu/Fe interface (thickness of the sampling volume : 0.5nm)*

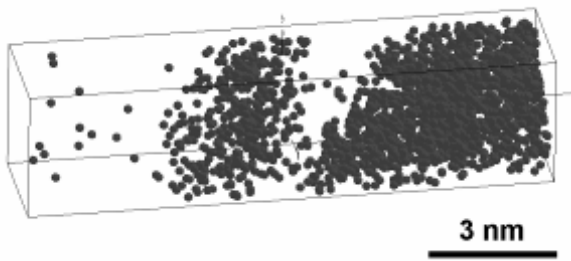
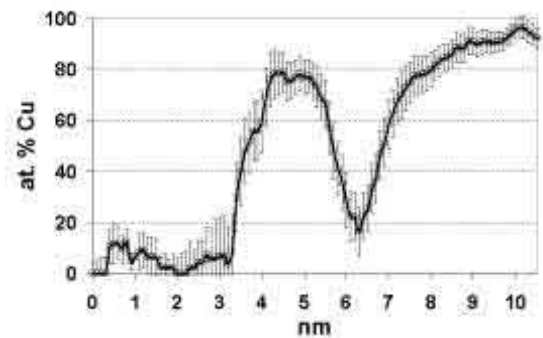

a)                                          b)

*Figure 7 : 3D-AP data set of the Cu-Fe composite after 5 turns by HPT, a) 3D reconstructed volume (3x3x12 nm³), only Cu atoms are displayed to exhibit nanoscaled Cu and Fe grains , b) Copper concentration profile across Cu/Fe interfaces (thickness of the sampling volume : 0.5nm).*

Another set of data was collected in a region with a 2nm thick Fe cluster (figure 7-a). The Cu concentration profile across the analysed volume exhibits Fe gradients and a strong Fe-Cu interdiffusion (figure 7-b). The Fe concentration in the copper phase is up to $20\pm3$ at.%. Thus, these data give the evidence of mechanical alloying of Fe and Cu during the HPT process but an homogenous supersaturated solution is not obtained. The driving force and the kinetics of this alloying reaction produced by severe plastic deformation is discussed in the following part.

## 4. Discussion

### *4.1 Decrease of the grain size during HPT*

The grain size of materials processed by HPT is usually in the submicron range. This is the result of both the elongation of grains and the formation of new high angle grain boundaries [1]. In this study, TEM observations of the Fe-Cu composite processed by HPT revealed two kinds of regions with different grain sizes :

i) Pure FCC copper regions which might be the tracks of former Cu tubes used to make the composite at the 2$^{nd}$ or 3$^{rd}$ step. They exhibit a grain size in a range of 100 to 200 nm which is consistent with experiments performed by other authors on pure Cu samples [8-11]. As reported by Valiev and co-authors, it is close to the minimum grain size achievable in severely deformed copper. They have shown indeed, that the grain size continuously decreases up to 5 turns of deformation, and then saturates at about 100nm [7]

ii) The BCC $\alpha$-Fe and FCC Cu mixed regions exhibit a smaller grain size in a range of 20 to 80 nm. This is much smaller than in single phase FCC copper [8-11] or BCC $\alpha$-Fe [4-7] processed by HPT. Thus, these data confirm the ability of multi-phase materials to produce structures with a grain size of several tens of nanometers through SPD [13-20, 24]. The size of Fe grains after HPT is however one order of magnitude larger than estimated on the basis of the macroscopic shear (about 50 nm instead of 5nm). This indicates that filaments are not continuously sheared during the torsion. This feature might be attributed to the curled shape of the original Fe filaments, to grain boundary sliding [32] or to a larger deformation rate of the Cu softer phase.

*4.2 Cu-Fe mixing*

*Supersaturated solid solution*

Although the mutual solubility of Fe and Cu is very small under equilibrium conditions (less than 0.1at.% at room temperature) [33], our 3D-AP data show that mechanical alloying of Cu and Fe might occur as a result of SPD. Such feature has been widely reported for ball-milled powders [16,25,26,34] but in the present study it might involved different mechanisms since it is the result of continuous shearing during the HPT process.

In this study, SEM back-scattered images (figure 3), FIM pictures (figure 5) and 3D-AP data (figures 6 and 7) show that the distribution of Fe atoms is not homogenous. The interdiffusion of Fe and Cu is deeper where the grain size is smaller. Since XRD and SAED patterns (figures 2 and 4) reveal only the BCC α-Fe and the FCC Cu phases, one may conclude that SPD had locally led to the formation of supersaturated solutions. Following the Vegard's law, the slight decrease of the FCC-Cu lattice parameter confirms this opinion. This feature is however not consistent with Cu-Fe supersaturated solid solutions produced by high energy ball-milling. They usually exhibit an increase of the FCC-Cu lattice parameter which has been attributed to elastic strains and magnetovolume effects [25].

*Driving force*

The HPT process dramatically reduces the grain size, so that some Fe filaments are transformed into nanometer scaled fragments (figure 7). The effect of the capillary pressure on Fe fragments in such a system could lead to significant changes in the phase diagram [35]. Yavari and co-authors have estimated that particles with a diameter of 2nm might spontaneously dissolved in the Cu matrix [16]. Thus they believe that during the ball-milling process needle-shaped particles are formed with tip radii small enough to reach a significant driving force for the dissolution.

Our 3D-AP data show the influence of the size of iron fragments on the amount of Fe in the copper phase. Figure 6 shows an iron fragment with a diameter of about 10 nm. In that case, the Fe concentration in the surrounding copper matrix reaches 4at.%. In the case of a 2nm thick iron fragment (figure 7), a much larger amount of Fe is detected in the copper phase (up to 20 at.%). Thus, the smaller the radius of iron fragments, the higher the Fe supersaturation of the Cu phase. This behaviour is consistent with the model proposed by Yavari and co-authors. The strong reduction of the grain size and capillary pressures might be therefore the thermodynamic driving force leading to the mechanical alloying of Fe and Cu.

*Kinetics*

The samples were processed by HPT at room temperature and due to the high thermal conductivity of anvils, the temperature of the sample remains below 50°C [36]. In such conditions, atomic mobilities of Cu and Fe are obviously too low to give rise to a significant interdiffusion [37]. So, the HPT process must promote at least one diffusion mechanism. Several features could enhance the atomic mobility. The first one is the atomic transport along grain boundaries (GB). Since, the proportion of GB is huge in such nanoscaled systems and while they move during the HPT process, they may promote the interdiffusion as suggested by Straumal and co-authors [36]. But these authors point out that GB diffusion decreases by several orders of magnitude in HPT condition due to the high pressure applied to the sample. The second mechanism is the pipe diffusion along dislocations [55]. As a matter of fact, SPD produces a high dislocation density. However, these defects are mostly located along GB [1] and since the strain rate is about 1 $s^{-1}$ during HPT deformation, they may not be able to drag solute atoms while they move.

This discussion will focus on the third possible mechanism : an increase of the mobile vacancy concentration. Indeed, they are many examples in the literature of phase transformations induced by SPD and most of them seem to indicate that additional vacancies are produced during the deformation [36, 38-41]. In their theory on "driven alloys" Martin and Bellon even suggest that forced atomic jumps in ball-milled powders are due to additional vacancies which concentration is proportional to the milling intensity [42]. Such vacancies induced by SPD have been revealed by Van Petegem and co-authors using positron lifetime spectroscopy of pure nickel samples processed by HPT but they do not provide any estimate of the vacancy concentration [43]. Kiritani and co-authors have performed TEM investigations of point defects (vacancy clusters) produced in thin films deformed at high strain rate [44,45]. They measured in gold samples a vacancy concentration up to $10^{-4}$ which is close to the concentration near the melting point. In copper samples deformed at a strain rate of 1 $s^{-1}$ the vacancy concentration was estimated up to $10^{-5}$ [44]. These additional vacancies would result from the recombination of dislocations [47,48], but under high internal stresses and where the nucleation of dislocations is difficult (like in nanoscaled systems subjected to HPT), they might result of a dislocation free plastic deformation [44,45].

In order to estimate the interdiffusion of Fe and Cu atoms, one has to estimate both the vacancy concentration during the HPT process, and the mobility of these vacancies in HPT conditions.

*i) Vacancy concentration during the HPT process*

It is interesting to note that the vacancy formation energy $E_f$ is significantly lowered in nanoscaled particles. Thus, Qi and co-authors have calculated that the vacancy formation energy is lowered by 10% in a gold particle with a diameter of 10 nm comparing to the bulk [49]. As demonstrated by Sato and co-authors, the internal stress is another important parameter [50]. During the HPT process, the compressive strain $\varepsilon_c$ in copper is about 4% ($\varepsilon_c$ = P / E , where P = 5GPa is the applied pressure and E = 120 MPa is the copper elastic modulus). Such a strain might be enough to promote the formation of excess vacancies. As estimated by Sato and co-authors, a compressive strain of 5% (respectively 10%) could decrease the vacancy formation energy from 1.20 eV down to 1.12 eV (respectively 0.74eV) [50].

The vacancy concentration under thermal equilibrium is given by [46] :

$$C_v = \exp(S_f / k_B) \exp(-E_f / k_B T) \quad (2)$$

where $S_f$ is the vacancy formation entropy, $E_f$ the vacancy formation energy, T the temperature and k the Boltzmann constant. The numerical calculation for copper ($S_f = 1.5\ k_B$ and $H_f = 1.2eV$ [46]) gives at room temperature $C_v \approx 3\ 10^{-20}$. If the vacancy formation energy is lowered by 20%, this concentration and thus the diffusion coefficient increases by four orders of magnitude. This estimate is much lower than experimental value up to $10^{-5}$ provided by other authors [44,45,51,52] and it gives the evidence that additional vacancies are directly provided by the plastic deformation. The experimental data of Kiritani and co-authors show that at a strain rate of about 1 s$^{-1}$ during about 1 s, the vacancy concentration in copper at room temperature rises up to $10^{-5}$ leading to vacancy clusters formation [44]. Since vacancies could be eliminated on crystal defects, the vacancy production rate is at least $10^{-5}$ s$^{-1}$. During the HPT process, a similar rate might be reached so that new vacancies are continuously produced and eliminated on crystal defects (dislocations, grain boundaries and interstitial atoms) [53]. It seems therefore reasonable to estimate that the mobile vacancy concentration is increased by 15 orders of magnitude during the HPT process.

## ii) Vacancy mobility in HPT conditions

If the vacancy migration entropy is neglected, the diffusion coefficient writes as [54] :

$$D = A \, \nu \, a^2 \, C_v \, \exp(-E_m / k_B T) \quad (3)$$

Where A is a geometrical factor depending on the crystal structure, ν the "attempt frequency", a the lattice parameter and $E_m$ the vacancy migration energy. In the copper FCC phase, $E_m \approx$ 0.7 eV [46], but Sato and co-authors have calculated that under a compressive strain of 5% the vacancy migration energy might be lowered by 30% [29]. Thus, in the specific conditions of the HPT process this would lead to an increase of four orders of magnitude of the diffusion coefficient.

## iii) Interdiffusion of Cu and Fe atoms

The diffusion coefficient of Fe in the copper phase at 30°C (lower temperature achieved during the HPT process) is $D_{Fe} \approx 5 \, 10^{-35} \, cm^2 \, s^{-1}$ [37]. An increase of 19 orders of magnitude (15 for the vacancy concentration and 4 for the vacancy mobility) of this diffusion coefficient could give rise to a significant increase of the Fe concentration in the copper phase. In equilibrium conditions, such a high atomic mobility is indeed reach at a temperature of about 710K. The diffusion length L of Fe atoms in the copper phase as a function of time t writes as [54] :

$$L = (6 \, D_{Fe} \, t)^{1/2} \quad (4)$$

The sample was processed by HPT during 1500 seconds (5 turns at a rate of 0.2 turn per minute), providing a diffusion length of about 20nm which is of the same order of magnitude than the grain size. Thus, additional vacancies induced by SPD are though to explain the mechanical alloying of Fe and Cu which was pointed out by 3D atom probe data (figures 6 and 7).

## 5. Conclusion

1- HPT deformation of a Cu-Fe filamentary composite leads to a layered nanostructure. The grain size of copper regions is in a range of 100 to 200 nm while regions containing the $\alpha$-Fe phase exhibit a much smaller grain size in a range of 20 to 80 nm. This confirms the ability of multi-phase materials to produce structures with a grain size of several tens of nanometers thanks to SPD.

2- TEM and X-ray diffraction data show that both the FCC copper and the BCC $\alpha$-Fe phases remain after the HPT deformation. But 3D-Atom Probe data show that locally there is a strong interdiffusion of Cu and Fe.

3- Fe nanoclusters are dissolved during SPD, and the smaller the clusters the more they dissolved in the copper matrix. Thus, the driving force for the dissolution is attributed to capillary forces.

4- HPT deformation would provide a vacancy production rate of $10^{-5}$ $s^{-1}$. These additional mobile vacancies have a low migration energy due to the high pressure applied to the specimen and are continuously eliminated on defects like dislocations and GB. This high level of vacancy increases the diffusion coefficient so that the diffusion length is of the same order of magnitude than the grain size. Thus, vacancies induced by SPD allow the mechanical alloying of Cu and Fe.